\documentstyle{article}
\begin{document}

\begin{center}

{\bf {\Large WHICH RADIUS FOR THE SUN? }}\\
\vspace{0.2 cm}
 {V. Castellani,$^1$ S. Degl'Innocenti,$^{1,2}$ G.Fiorentini $^{2,3}$} \\

\end{center}
\vspace{0.5 cm}
\noindent
  $^1$ Dipartimento di Fisica, Universit\'a di Pisa, piazza Torricelli 2, I-56126 Pisa, Italy \\
  $^2$ Istituto Nazionale di Fisica Nucleare, Sezione di Ferrara, via Paradiso 12,
 I-44100 Ferrara, Italy\\
  $^3$ Dipartimento di Fisica, Universit\'a di Ferrara, via Paradiso 12,
 I-44100 Ferrara, Italy\\

\vspace{1 cm}
%\date{Accepted 1988 December 15.
%      Received 1988 December 14;
%      in original form 1988 October 11}

%\pubyear{1994}

%\maketitle

%\label{firstpage}

\begin{abstract}

The high accuracy reached by solar limb observations, by helioseismic
measurements and by Standard Solar Models (SSMs)
calculations suggests that general
relativity corrections are included when discussing
the solar radius. The Allen value (R$_{\odot}$ = 695.99 $\pm$ 0.07 Mm)
has to be reduced by 1.5 Km. This correction, which is small as compared
with present accuracy, should be kept in mind for future more precise
measurements and/or calculations.

\end{abstract}

\vspace{0.8 cm}
%\section{}

Following a well established procedure, the observation of the 
solar disc combined with precise estimates of the astronomical unit 
has early  produced high quality evaluations of the solar radius.
According to  Allen (1976) one finds: 
\begin{equation}
R_{\odot}  = 695.99 \pm 0.07 ~{\rm Mm};   \hspace{0.3 cm}  ( \Delta R/ R )_{\odot} = \pm 1 \cdot 10^{-4}.
\end{equation}
As recently reviewed by Schou et al. (1997) and Brown and Christensen-Dalsgaard (1998),
more recent estimates appear in substantial agreement with such a value. 
Thus, for a long time this value has been used to constrain Standard Solar Models   
with a precision largely exceeding the common requirements for
stellar evolutionary models.

However, in more recent time helioseismology (HeS) has proved to
be able to test the solar structure to an extraordinary degree of
precision, comparable with the error quoted above, see e g. Basu and Christensen-Dalsgaard (1997),
Bahcall, Basu and Kumar (1997), Degl'Innocenti et al. (1997), Dziembowski (1996), Dziembowski et al. (1994).
As a consequence, recent works (Schou et al., 1997, Basu, 1998, and Antia, 1998)
discussed the possibility that helioseismology indicates a  slightly smaller 
radius for the Sun, decreasing the above quoted value by a quantity 
\begin{equation}
\delta R_{HeS} \approx 0.3 ~{\rm Mm};  \hspace{0.3 cm}   ( \delta R/ R )_{HeS} \approx 4 \cdot 10^{-4}                 
\end{equation}
which, though rather small, is definitely beyond the formal error
for Allen's value.

 The issue has been recently discussed by Brown and
 Christensen-Dalsgaard (1998). They reanalyzed 
the already known inconsistency between 
the observational radius, R$_{\odot}$, 
defined as the radial distance to the inflection point of the limb intensity profile,
and the photospheric radius R$_{ph}$, adopted by theoretical models, which is
the point at the basis of the atmosphere with a temperature T equal to the
star effective temperature, i.e.,  the point such
that $R^2_{ph} T^4$=L$_{\odot} / (4 \pi \sigma)$ (see e.g. Wittmann, 1974,). According
 to such a definition, the authors, 
by using updated model atmospheres, find that an appropriate SSM should
be calculated for a radius R$_{ph}$ smaller with respect to R$_{\odot}$
by an amount:

\begin{equation}
\delta R_{ph} = 0.5 ~{\rm Mm};  \hspace{0.3 cm}          ( \delta R/ R )_{ph} = 7 \cdot 10^{-4}.                          
\end{equation}

Note that such a correction appears quite negligible in current evolutionary
evaluations, achieving a relevance only in connection with the
stringent constraint from HeS. As a matter of fact, in the past, 
SSMs (as given, e.g., by Bahcall and Pinsonneault 1995,
 Christensen-Dalsgaard et al. 1996, Ciacio et al. 1996) have been calculated 
for a radius larger than R$_{ph}$ but, in the meantime,
the suggested correction would push the radius 0.2 {\rm Mm} below the
value suggested by HeS. 

Without entering into a discussion on that matter, in this note we discuss the
effect of gravitation on the apparent solar radius.
The maximum impact parameter b of photons reaching Earth corresponds
to trajectories tangent to the solar surface, of radius R$_0$.
The relationship between b and R$_0$ is easily derived by using the conservation laws
of energy and of angular momentum. In the weak field approximation, a
photon with energy h$ \nu _0$ at the Sun surface when at large distance has energy given by:
\begin{equation}
 h \nu = h \nu _0 (1 - \frac{GM_{\odot}}{ R_0 c^2}).
\end{equation}

Angular momentum conservation gives:
\begin{equation}
\frac{ h \nu b}{c} = \frac{h \nu _0 R_o}{c}.
\end{equation}
 The above equations immediately give:
\begin{equation}
b= \frac{R_0 \nu _0}{ \nu} \approx R_0 + \frac{GM_{\odot}}{c^2}.
\end{equation}
 
In order to get the true solar radius  R$_0$ from the apparent solar radius 
R$_{\odot}$  = D sin$\alpha _{app}$ = b one has to subtract a quantity:

\begin{equation}
 \delta R_{GR} = \frac{GM_{\odot}}{c^2} = 1.5 ~{\rm Km}; \hspace{0.3 cm}  ( \delta R/ R )_{GR} \approx  2 \cdot 10^{-6}. 
\end{equation}

This correction is small in comparison with the present accuracy of solar observations and/or
helioseismic determination. However, it should be kept in mind for the future.\\

\begin{Large}
{\bf Acknowledgments}\\
\end{Large}

We appreciate discussions with J.N. Bahcall, S. Basu, E. Iacopini and W. Dziembowski.
We are grateful to T. Brown, J. Christensen-Dalsgaard and G. Rybicki for 
pointing to our attention an important mistake in the previous version of
this note.

\end{document}